# Experimental Study of Resistive Bistability in Metal Oxide Junctions

Zhongkui Tan[a)], Vijay Patel, Konstantin K. Likharev

*Department of Physics and Astronomy, Stony Brook University, Stony Brook, NY 11794-3800*;

Dong Su, Yimei Zhu

*Center for Functional Nanomaterials, Brookhaven National Laboratory, Upton, NY 11973-5000*

*Abstract*

We have studied resistive bistability (memory) effects in junctions based on metal oxides, with a focus on sample-to-sample reproducibility which is necessary for the use of such junctions as crosspoint devices of hybrid CMOS/nanoelectronic circuits. Few-nm-thick layers of $NbO_x$, $CuO_x$ and $TiO_x$ have been formed by thermal and plasma oxidation, at various deposition and oxidation conditions, both with or without rapid thermal post-annealing (RTA). The resistive bistability effect has been observed for all these materials, with particularly high endurance (over $10^3$ switching cycles) obtained for single-layer $TiO_2$ junctions, and the best reproducibility reached for multi-layer junctions of the same material. Fabrication optimization has allowed us to improve the OFF/ON resistance ratio to about $10^3$, but the sample-to-sample reproducibility is so far lower than that required for large scale integration.

---

a) Author to whom correspondence should be addressed; email: tanzhongkui@gmail.com

## I. INTRODUCTION

The exponential, "Moore's Law" progress of semiconductor integrated circuit technology[1] will face though challenges in just a few years, mostly due the necessity of sub-nanometer accuracy of field-effect transistor feature definition – see, e.g., Ref. 2. One of the most attractive options to bypass these problems and extend the Moore's Law by estimated 10 to 15 years[3] is to use hybrid CMOS/nanoelectronic circuits. The simplest, and apparently the most promising version of such circuit[2-9] is a usual CMOS stack, augmented with a back-end nanoelectronic crossbar,[10] with similar nanodevices formed at every crosspoint.

The first suggestions[4,5] of the hybrids required using complex, three-terminal devices, whose implementation is still well beyond experimental reach. However, it was soon realized that the circuits may function very effectively even if the crosspoint device is just a diode (connecting two nanowires, one from each crossbar layer) with resistive bistability (Fig. 1a).[11] Experimental observation of such bistability in many materials (including metal oxides and sulphides, amorphous silicon, organic layers with and without embedded metallic grains, and molecular self-assembled monolayers) have been reported in numerous publications starting at least from the 1960s – see an early review[13] and recent reviews.[9,14,15] Moreover, these studies have led to a virtual consensus that the resistive bistability, at least in metal-oxide and amorphous-silicon junctions,[16] is due to the reversible formation and dissolution of one or few highly conducting spots (sometimes called "filaments"), due to field-induced drift of ions (depending on the particular material, either anions or cations) through the amorphous matrix of the layer – see Fig. 1b-d.[9,14,15]

Because of this atomic-scale mechanism, the most critical feature of the bistable junctions, especially in the view of their possible applications in VLSI circuits, is the device-to-device reproducibility. However, most publications do not give any quantitative information about the achieved reproducibility. We are aware of just a few exceptions:

(i) A Samsung group has published[17] histograms of ON and OFF resistances of junctions of an unspecified metal oxide, with two substantially different areas, 0.2 and 0.0025 $\mu m^2$. In both cases, the statistical distributions of ON and OFF resistances form relatively narrow peaks (below one order of magnitude wide) which are well separated, by approximately factors 300 and 30, respectively. Unfortunately, no statistics has been given for switching threshold voltages $V_t$ and $V'_t$ (Fig. 1a), the bistability parameters most critical for applications.[9]

(ii) A Spansion team has presented[18] somewhat less impressive current histograms for their 0.18×0.18 $\mu m^2$ $CuO_x$ junctions with a 12-nm oxide layer; still, the ON and OFF current values are separated by a gap of at least one order of magnitude wide. Again, no switching threshold statistics have been reported.

(iii) A University of Michigan at Ann Arbor group did present[19] a histogram of one of switching thresholds ($V_t$ in Fig. 1a) of their 50×50 $nm^2$ junctions with an 80-nm-thick amorphous-silicon layer. The histogram features a very narrow (~10%) peak, at apparently much larger split between average values of $V'_t$ and $V_t$.

(iv) A collaboration of the Chinese Institute for Microelectronics and University at Albany have reported[20] a narrow but still clean separation of ~30%-wide histogram peaks for $V'_t$ and $V_t$, in 0.5×0.5 $\mu m^2$ junctions consisting of three sequentially deposited $ZrO_2$/Cu bilayers, with a thickness of 20+3 nm each.

(v) Finally, very recently, a group from Gwangju, Korea reported[21] a huge (~4 orders-of-magnitude) gap between the threshold histogram peaks (each less than an order-of-magnitude wide) in 0.5×0.5 $\mu m^2$ junctions based on ~70 nm thick layers of a polyfluorene-derivative polymer.[22]

However, even these publications report only the apparently best results, and do not describe how sensitive they have been to variations of the fabrication conditions. The goal of this work has been to explore bistability effects in junctions based on oxides of Cu and Ti, which looked most promising from literature data (plus Nb which was a legacy metal for our laboratory), within a broad range of fabrication and post-processing conditions and procedures of the electric "formation" of the devices. In contrast to virtually all other publications in this field, we present experimental data on device reproducibility (and also other important properties such as OFF/ON conductance ratio and switching endurance), regardless on whether they look favorable or unfavorable. We hope that these data will give important clues to other research groups in pursuit of the important goal of integrable nanodevice development.

## II. FABRICATION AND EXPERIMENTAL PROCEDURES

Most metal-oxide layers of our junctions were fabricated by either rf plasma or thermal oxidation of a thin metallic layer (or layers) on 2" thermally-oxidized silicon wafers, at ambient temperature. The fabrication procedures of two types were used.

<u>(i) Vacuum-break process</u> (wafers VJCuOx3, 4, 5, 6, 7 and VJTiOx1, 2, 3):

A layer of metal base electrode (50-100nm) was first deposited by electron beam evaporation, at a 0.05 to 0.08 nm/s rate, in a ~$5\times10^{-7}$ Torr vacuum. After a vacuum break, the sample was rapidly transferred to a sputtering chamber with base vacuum of 2 to $3\times10^{-7}$ Torr. After pre-cleaning in an rf Ar plasma for a time sufficient to remove ~5 nm of the base electrode, either the thermal oxidation (at 100 Torr pressure of dry oxygen, for 10 to 40 minutes), or rf plasma oxidation (at 10 to 300 W rf power, at 15 to 30 mTorr $O_2$ pressure, for 10 minutes) was performed. The few-nm-thick oxide layer was then sealed by a 100-nm-thick Nb counter-electrode film, dc-sputtered at the rate close to 2 nm/s.

(ii) In-situ processes (all other wafers listed below):

The whole junction structure was fabricated in a single vacuum system equipped either for sputtering (for $NbO_x$ and $CuO_x$) or e-beam evaporation (for $TiO_x$). The in-situ process has enabled us to provide larger variety of metal electrodes (see Tables 1-3 below for details) and cleaner interface between the metal and metal-oxide layers. For $NbO_x$ devices, a 50-nm-thick Al wiring level was first dc-sputtered at 0.5 nm/s, followed by a 10-nm-thick Nb base layer. For $CuO_x$ samples, the substrate was pre-coated with a 5-nm Cr adhesion layer, followed by dc-sputtering, at a rate of ~2 nm/s, of a 150-nm-thick Cu base electrode. Following the surface oxidation, stacks of both types were completed by dc-sputtering of ~100-nm-thick Nb counter-electrodes at a rate ~ 2 nm/s.

For $TiO_x$-based junctions, the deposition of a similar Cr adhesion layer was followed by e-beam evaporation of 50 to 100 nm Pt wiring layer and its lift-off patterning. Then the wafer was cleaned from any resist and chemical residue in an oxygen rf plasma asher and moved into the e-beam chamber, where it was cleaned again in rf Ar plasma as described above, before the deposition of the titanium layers. For Ti/$TiO_x$/Ti devices, a 50-100 nm thick Ti electrode was e-beam evaporated at ~0.05 nm/s, followed by e-beam evaporation of $TiO_2$ from a stoichiometric target. For Pt/$TiO_x$/Ti type devices, a very thin (1.5 nm) layer of Ti was evaporated on the Pt base, and then exposed to oxygen-enriched rf plasma to completely oxidize the layer. For multi-layer $TiO_x$ junctions, this process was repeated several times. In both cases, the oxide layer was sealed by e-beam evaporation of a 100 nm thick Ti counter-electrode, as described above.

Figure 2 shows an annular dark-field scanning transmission electron microscopy (ADF STEM) image of one of our multi-layer samples (wafer VJTiOx8). It shows sharp, clean, and relatively smooth interfaces between the layers.

After the stack had been fabricated, it was patterned to define 18 junctions of areas 3×3, 30×30, and 300×300 $\mu m^2$, with appropriate wiring and contact pads, on each 5×5 $mm^2$ chip. For that, Nb and Ti electrode patterns were defined by the reactive ion etching in $SF_6$ gas using a PMMA etch mask patterned with UV lithography. Other metals (viz. Al and Cu) were patterned by back-sputtering in Ar rf-plasma again using PMMA etch masks. A 150 nm thick rf-sputtered quartz layer was used as for insulation and patterned with a self-aligned-liftoff process using the junction layer etch mask. A final wiring layer of 200-nm-thick, dc-sputtered Nb was patterned via lift-off.

After initial junction testing, several chips from most fabricated wafers were subjected to rapid thermal post-annealing (RTA) in Ar flow, at temperatures from 200 to 800ºC, for 30 to 180 seconds. (For particular values, see Tables 1-3.)

Since typical junction resistances were in excess of $10^2$ Ω, i.e. larger than that of thin-film on-chip wiring, their electrical characterization was performed by simple two-terminal *I-V* measurements. For the initial formation of the ON-state (i.e. Fig1b-d), applied voltage was increased, with current externally limited to a certain value, typically of a few mA. (The so-called "current compliance".)

Voltage sweeps were performed at a speed of 1 to 100 mV/s. In the junctions exhibiting resistive bistability, the typical OFF→ON switching time was less than 10 μs (our measurement technique limit), while the typical ON→OFF switching took much more time, in the range of milliseconds. For quantitative characterization of ON and OFF states, the corresponding resistances $R_{ON}$ and $R_{OFF}$ were measured at low bias voltage (~50 mV). The resistive bistability cycle could be typically repeated several (*N*) times, usually followed by a hard breakdown to an irreversible state with a very low resistance.

ON/OFF switching statistics was recorded for all devices which exhibited the bistability. The "yield" listed in Tables 1-3, was defined as ratio of number of samples with resistive bistability behavior to the overall number of all samples without evident microshorts.

## III. RESULTS: $NbO_x$

Our laboratory has long experience of fabrication of high-quality thin-film structures based on niobium, so that in light of several prior publications[27,28] reporting the resistive bistability in junctions based on oxides of that material, it was natural for us to start our experiments with such devices. Table 1 summarizes the major parameters and properties of our $Nb/NbO_x/Nb$ junctions.

Our initial attempt of forming the oxide layer by simple thermal oxidation in dry oxygen (wafer VJNbOx1), has produced a very low yield. The post-annealing did not help much.

The transfer to plasma oxidation, at modest rf power (wafers VJNbOx2 and 3), has not increased the yield of as-grown junctions. However, such devices have benefited more from the RTA (Fig. 3), with the yield clearly growing with RTA temperature until it reaches ~500°C. Unfortunately, at approximately the same temperature, the OFF/ON resistance ratio starts to drop rapidly (Fig. 3b). Moreover, the switching endurance of such junctions, characterized by the number $N$ of ON/OFF switching cycles (like that shown in Fig. 3a) was low, with the typical $N$ of the order of 10 or so.

An attempt to improve the situation by the further increase of rf plasma power (wafer VJNbOx4) has given junctions with typical Schottky-barrier *I-V* curves, without observable hysteresis.

Since by that time, we had reached more promising results with $CuO_x$ devices, we decided not to pursue the niobium oxide option any longer.

## IV. RESULTS: $CuO_x$

Experiments with copper oxides (Table 2) were also started with thermal oxidation – see wafer VJCuOx3, which gave similarly poor yield.

The transfer to plasma oxidation, at modest rf power (10 to 100 W) has helped a lot, especially when accompanied by the RTA, with the average yield rising to ~50% (Fig. 4) for wafer VJCuOx7. Unfortunately, just like in the case of $NbO_x$, the yield rise is accompanied by a sharp drop of the OFF/ON resistance ratio.

Encouraged by prior work,[29] we have explored the option of very high plasma power combined with a higher oxygen pressure (wafers VJCuOx13, 15 and 17). Together with an RTA at 400°C, this has led to an improvement of the resistance ratio, but the yield has dropped.

In addition, the switching endurance for all copper-oxide junctions was rather low, with the number $N$ of cycles not exceeding 20 or so.

## V. RESULTS: $TiO_x$

In the view of recent encouraging publications,[24,26,30-33] the main focus of our work has moved to devices with titanium oxide interlayer(s) – see Table 3. Just as in the case of other two oxides, we have started with the simplest option of thermal oxidation (wafer VJTiOx1), just to get equally poor results.

Our attempts to use a completely different way of $TiO_x$ formation, by its evaporation deposition from a stoichiometric $TiO_2$ target (wafers VJTiOx4, 7 and 13) has also produced apparent *I-V* hysteresis loops, but they were very sensitive to temperature and the voltage sweep rate. A further study has shown considerable current change was happening even at fixed dc bias voltage, i.e. the measured states were not stable in time, putting in question the whole body of previously recorded data.

The transfer to plasma oxidation of the base titanium electrode, accompanied by post-annealing at relatively high temperatures (e.g., 700°C), has immediately improved the picture, with good junction yield reaching 50% - see Fig. 5a,b. The switching endurance was also improved to $N \sim 10^3$ (Fig. 5c), while the resistance ratio was not too impressive (see the rows for VJTiOx2 and 3 in Table 3), but acceptable for some applications.[9] Further attempts at a higher rf power and RTA temperatures did not help to improve resistance ratio, so that other fabrication methods were clearly needed.

In hope to improve the results even further, and inspired by recent publication,[31] we have explored in detail the option of several sequential cycles, each consisting of deposition of a very thin (1.5 nm) Ti layer, followed by its plasma oxidation (wafers VJTiOx6, 8, 9, 12, 14, 16, and 17). Such thin individual layers are hardly continuous (as partly confirmed by their HR TEM images like the one shown in Fig. 2), and their sequential deposition and thorough oxidation are just a good way to produce relatively thick, virtually uniform layer of $TiO_x$. For example, Fig. 2 shows the ADF-STEM image of a sample from wafer VJTiOx8, with 5 layers forming 13 nm of oxide. A detailed electron energy loss spectroscopy (EELS) study has shown that through this layer, the titanium-to-oxygen atomic ratio changed little, with the average value higher than 0.5, indicating some oxygen deficiency in comparison with the stoichiometric $TiO_2$. The study has also shown a certain fraction of Pt atoms in the oxide layer, gradually decreasing toward the counter-electrode, apparently due to some re-sputtering of the base electrode material in the oxidizing rf plasma, probably responsible for the layer non-uniformity visible in Fig. 2.

Such multi-cycle deposition gave us junctions with the best reproducibility to date, with ~70% junction yield, and close similarity of dc *I-V* curves of good junctions ($|V_t|$ and $|V'_t|$ ~1 V, see, e.g., Fig. 6). Unfortunately, the RTA, while increasing the resistance ratio to as high as $\sim 10^3$, and sustaining similarly high switching endurance, reduces the yield – see Fig. 7. Our attempts at more deposition-oxidation cycles to increase the oxide thickness (wafers VJTiOx9 and 17) gave a certain resistance ratio increase, but continuously reduced the good device yield, with much higher threshold voltages $V_t$ and $V'_t$.

The high yield obtained on our best wafer VJTiOx8, with 5 sequentially oxidized Ti layers, have allowed us to perform a more quantitative test of the sample-to-sample reproducibility, namely the measurements of switching threshold voltage statistics. The results are shown in Fig. 8. One can see a clear gap between the histogram peaks corresponding to $V_t$ and $V'_t$ .

VI. CONCLUSIONS

To summarize, we have explored the effect of resistive bistability in junctions with interlayers of three metal oxides, $NbO_x$, $CuO_x$, and $TiO_x$, formed by several techniques, within a broad range of fabrication and post-processing conditions – see Tables 1-3. The results indicate that the problem of reproducible resistive bistability is much less rosy than implied by most publications in the field. Namely, while the mere demonstration of the bistability is pretty straightforward with any of those oxides (and, by literature data, with many other materials), the implementation of device-to-device reproducibility, with high yield of good devices, is much harder.

So far, our best reproducibility results, with the yield close to 70%, and a clear separation of histogram peaks for two switching thresholds (Fig. 8), have been obtained for $TiO_x$ junctions with ~13 nm oxide layer formed by 5 sequential deposition-oxidation cycles, without post-annealing. While such reproducibility is on a par with the best results reported for metal oxide devices in the literature,[17,18,20] it is only sufficient for simple hybrid circuit demonstrations,[23-26] rather for real large-scale integration. We see the following reserves available for the further improvement of the reproducibility and other device parameters (such as the $R_{OFF}/R_{ON}$ ratio, switching endurance, and switching speed).

(i) Using junctions of much smaller area. Indeed, most interesting applications require much smaller (10-nm-scale) crosspoint devices,[9] and the apparent mechanism of bistability (see Fig. 1b-d and its discussion) may actually give more reproducible results for smaller junctions – the conclusion partly confirmed in Ref. 17.

(ii) Forming junctions with short voltage pulses (or their sequences), rather than the dc voltage used in our experiments. Such method may prevent local heating effects which may mask, or even reverse the field-induced ion drift.

(iii) Using different materials (such as amorphous-silicon[19] or polymer interlayers[22]) and/or different fabrication conditions.

It is our feeling that the task of reaching the ~90% device yield necessary for VLSI applications[9] is by no means hopeless, though it may require a large-scale industrial effort. We hope that our results will be useful for such effort.

ACKNOWLEDGMENTS

This work was supported by AFOSR. Fruitful discussions with W. Lu, J. E. Lukens and F. Miao, and technical assistance by S.-S. Chang and E. Monge are gratefully acknowledged.

# Tables

Table 1. Parameters and properties of $NbO_x$ samples

<table>
<tr><th>Wafers</th><th>Interlayer formation</th><th>Stack</th><th>RTA</th><th colspan="2">Bistability/Properties</th></tr>
<tr><td>VJNbOx1</td><td>thermal oxidation:<br>100 Torr $O_2$, 40 min</td><td>$Nb/NbO_x/Nb$</td><td>400˚C,<br>30 s</td><td>Y</td><td>yield <10%<br>$R_{OFF}/R_{ON}$ <10</td></tr>
<tr><td>VJNbOx2</td><td>plasma oxidation: 10 W,<br>15 mTorr $O_2$, 10 min</td><td rowspan="3">$Nb/NbO_x/Nb$</td><td rowspan="3">400 to<br>600˚C,<br>30 to<br>180 s</td><td>Y</td><td rowspan="2">yield <40%<br>$R_{OFF}/R_{ON}$ <10</td></tr>
<tr><td>VJNbOx3</td><td>plasma oxidation: 100 W,<br>15 mTorr $O_2$, 10 min</td><td>Y</td></tr>
<tr><td>VJNbOx4</td><td>plasma oxidation: 300 W,<br>5 Torr $O_2$, 10 min</td><td>N</td><td>Schottky barriers</td></tr>
</table>

Table 2. Parameters and properties of $CuO_x$ samples

<table>
<tr><th>Wafers</th><th>Interlayer formation</th><th>Stack</th><th>RTA</th><th colspan="2">Bistability/Properties</th></tr>
<tr><td>VJCuOx3</td><td rowspan="2">thermal oxidation:<br>100 Torr $O_2$, 40 min</td><td rowspan="2">$Cu/CuO_x/Nb$</td><td rowspan="2">400˚C,<br>180 s</td><td>Y</td><td rowspan="2">yield <5%<br>$R_{OFF}/R_{ON}$ <5</td></tr>
<tr><td>VJCuOx4</td><td>Y</td></tr>
<tr><td>VJCuOx5</td><td>plasma oxidation: 10 W,<br>15 mTorr $O_2$, 10 min</td><td rowspan="3">$Cu/CuO_x/Nb$</td><td rowspan="3">200 to<br>800˚C,<br><br>30 to<br>180 s</td><td>Y</td><td rowspan="3">yield ~50%<br>$R_{OFF}/R_{ON}$ ~2<br>(at 100 W;<br>RTA 800˚C,<br>30 s)</td></tr>
<tr><td>VJCuOx6</td><td>plasma oxidation: 50 W,<br>15 mTorr $O_2$, 10 min</td><td>Y</td></tr>
<tr><td>VJCuOx7</td><td>plasma oxidation: 100 W,<br>15 mTorr $O_2$, 10 min</td><td>Y</td></tr>
<tr><td>VJCuOx13</td><td>plasma oxidation: 100 W,<br>25 mTorr $O_2$, 10 min</td><td rowspan="3">$Cu/CuO_x/Nb$</td><td rowspan="3">400˚C,<br>30 s</td><td>Y</td><td rowspan="3">yield <20%<br>$R_{OFF}/R_{ON}$ ~10</td></tr>
<tr><td>VJCuOx15</td><td>plasma oxidation: 300 W,<br>25 mTorr $O_2$, 10 min</td><td>Y</td></tr>
<tr><td>VJCuOx17</td><td>plasma oxidation: 300 W,<br>25 mTorr $O_2$, 10 min</td><td>Y</td></tr>
</table>

Table 3. Parameters and properties of $TiO_x$ samples

| Wafer | Interlayer formation | Stack | RTA | Bistability/Properties | |
|---|---|---|---|---|---|
| VJTiOx1 | thermal oxidation: 100 Torr $O_2$, 40 min | Ti/$TiO_x$/Nb | 400˚C, 30 s | Y | yield <10% $R_{OFF}/R_{ON}$ <5 |
| VJTiOx2 | plasma oxidation: 50 W, 15 mTorr $O_2$, 10 min | Ti/$TiO_x$/Nb | 400 to 800˚C, 30 s | Y | yield ~50% $R_{OFF}/R_{ON}$ = 5 - 100 (at RTA at 700 ˚C, 30 s) |
| VJTiOx3 | plasma oxidation: 500 W, 5 Torr $O_2$, 10 min | | | Y | |
| VJTiOx4 | deposited $TiO_2$, thickness ≈ 15 nm | Ti/$TiO_x$/Ti | 400˚C, 30 s | N | metastable junctions (see the text) |
| VJTiOx7 | | Pt/$TiO_x$/Ti | | N | |
| VJTiOx13 | | | | N | |
| VJTiOx6 | plasma oxidation of 1.5 nm Ti (1 layer) | Pt/$TiO_x$/Ti | 400 to 700˚C, 30 s | Y | yield <30% $R_{OFF}/R_{ON}$ <30 |
| VJTiOx12 | | | | Y | |
| VJTiOx8 | plasma oxidation of 1.5 nm Ti (5 cycles) | | 200 to 700˚C, 30 s | Y | yield ~70% $R_{OFF}/R_{ON}$ = 30 - $10^3$ |
| VJTiOx14 | | | | Y | |
| VJTiOx16 | | | | Y | |
| VJTiOx17 | plasma oxidation of 1.5 nm Ti (7 cycles) | | 300˚C, 30 s | Y | yield <40% $R_{OFF}/R_{ON}$ = 50 - $10^3$ |
| VJTiOx9 | plasma oxidation of 1.5 nm Ti (10 cycles) | | 400˚C, 30 s | Y | yield <15% $R_{OFF}/R_{ON}$ >200 $V_t$ >5 V |

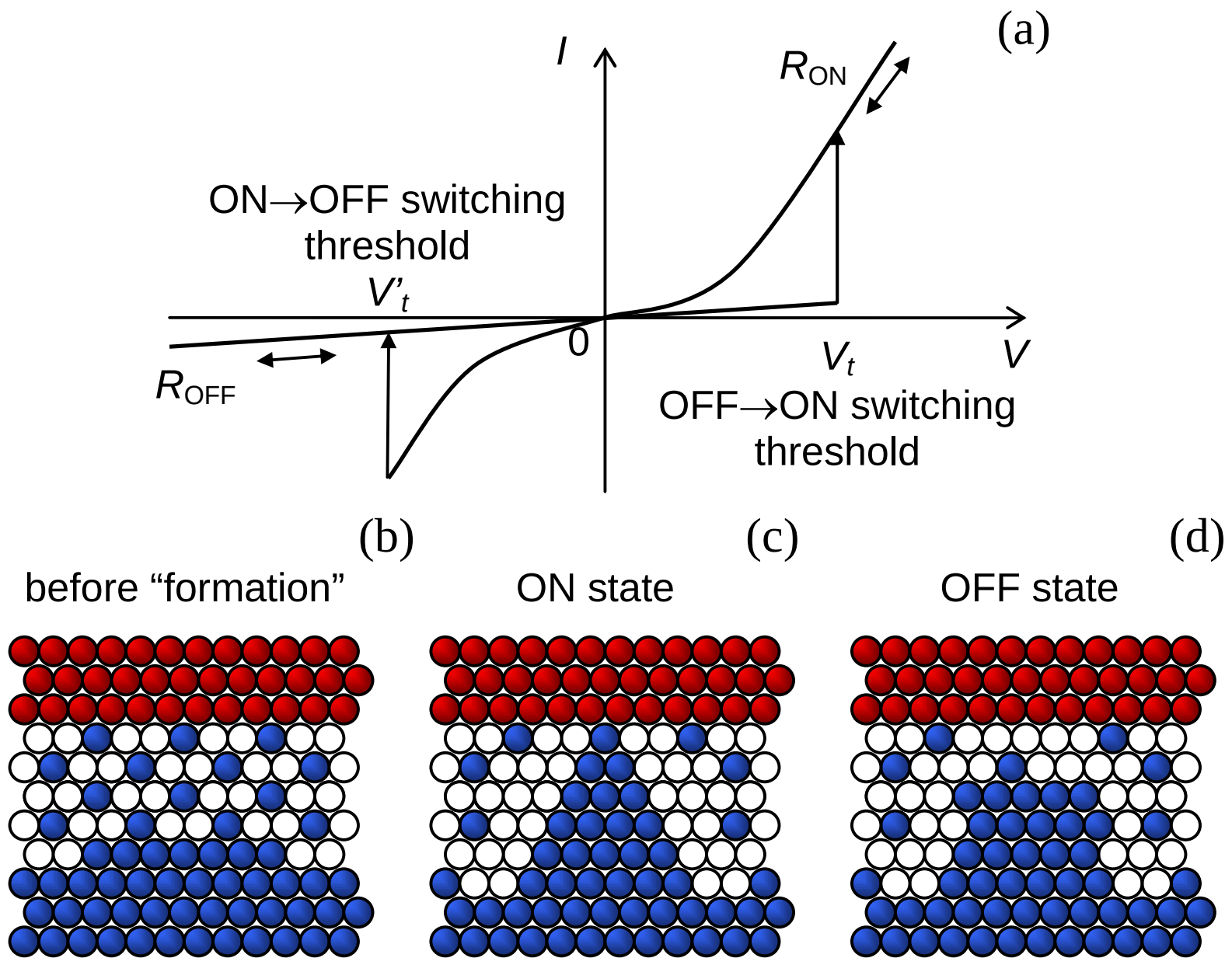

Fig. 1. Resistive bistability in metal oxides: (a) the dc *I*-*V* curve (schematically), and the parameter nomenclature; (b, c, d) a cartoon of the apparent bistability mechanism.

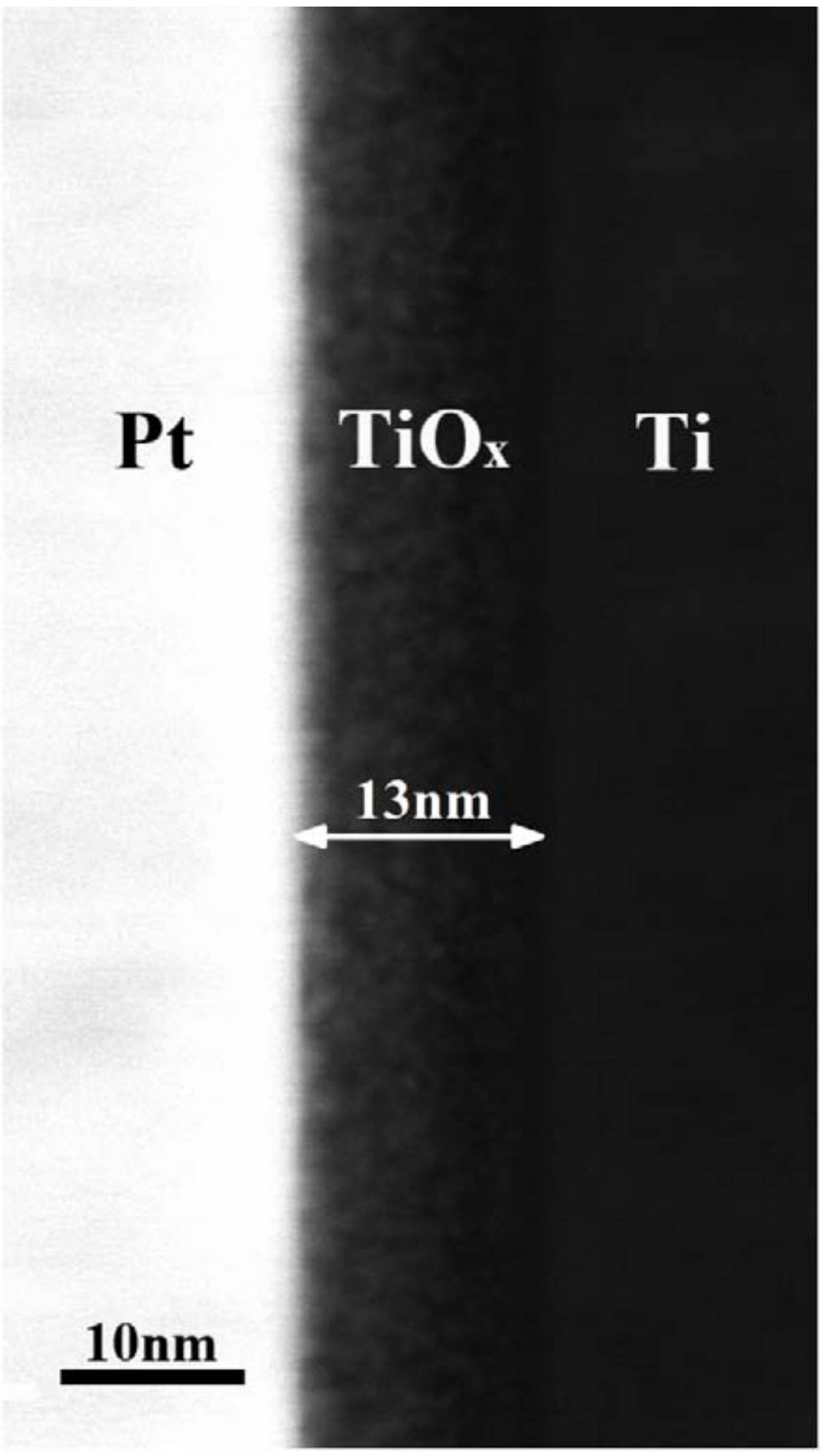

Fig. 2. An ADF-STEM image of a junction from wafer VJTiOx8.

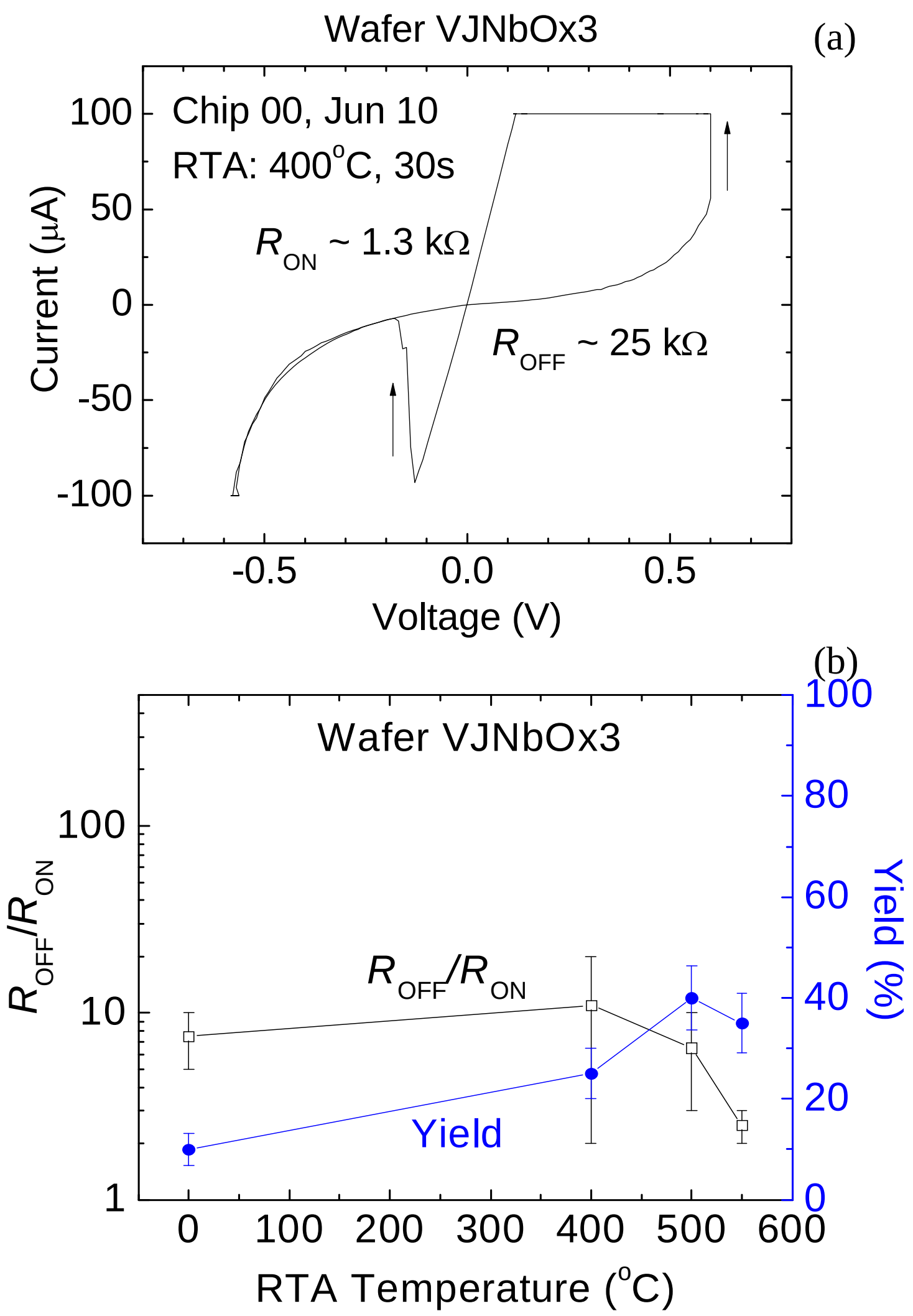

Fig. 3. (a) Typical dc *I-V* curve of a junction from wafer VJNbOx3, and (b) effect of temperature of a 30-second RTA on the OFF/ON resistance ratio and the yield of good devices from that wafer. The error bars correspond the r.m.s. scattering of the data among different samples. (The measurement accuracy was much better.)

(a)

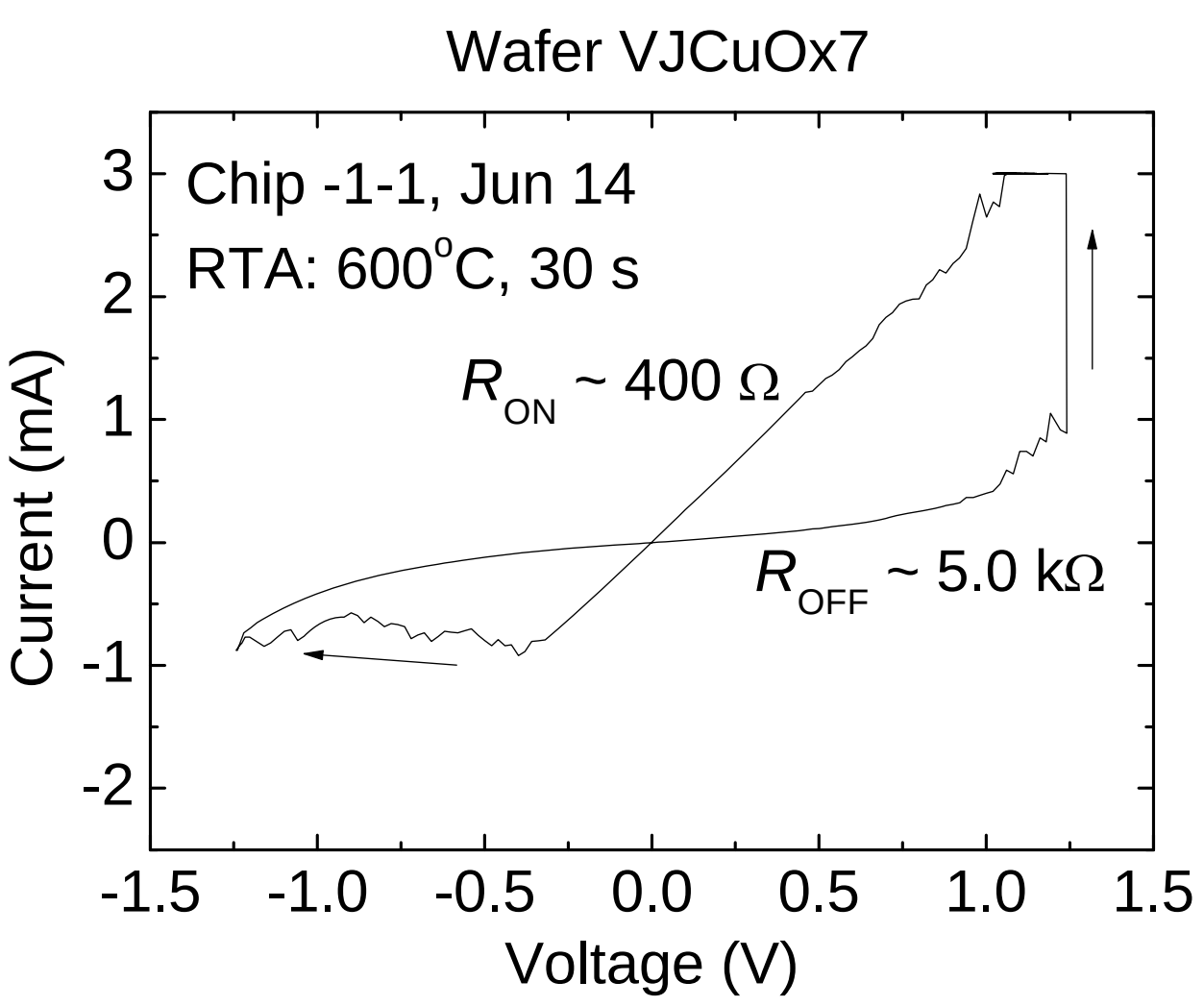

(b)

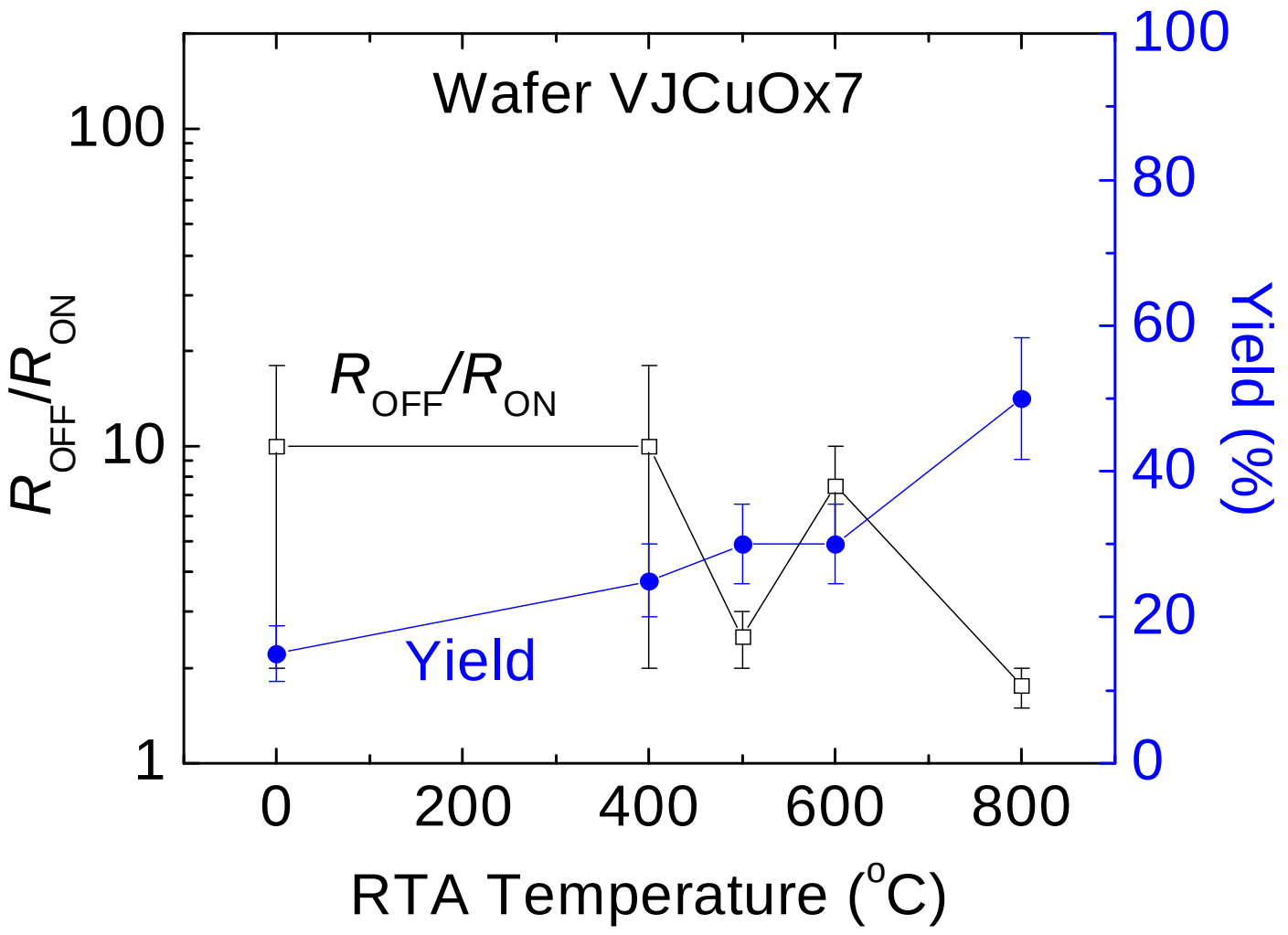

Fig. 4. The same as in Fig. 3, for wafer VJCuOx7.

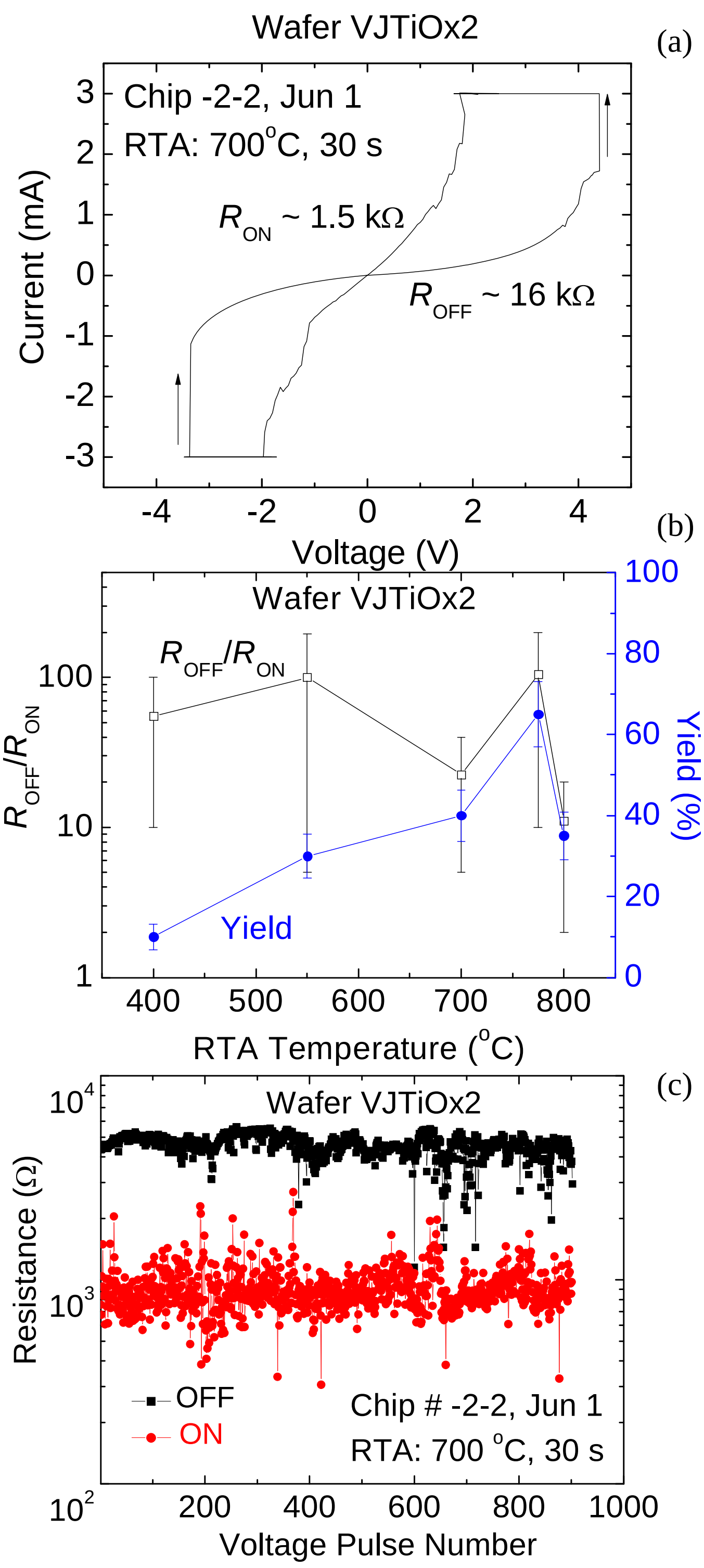

Fig. 5. (a) Typical dc *I-V* curve of a junction from wafer VJTiOx2, (b) effect of temperature of a 30-second RTA on the OFF/ON resistance ratio and yield of good devices from that wafer, and (c) results of the "endurance test" (repeated ON/OFF cycling) of one of the devices.

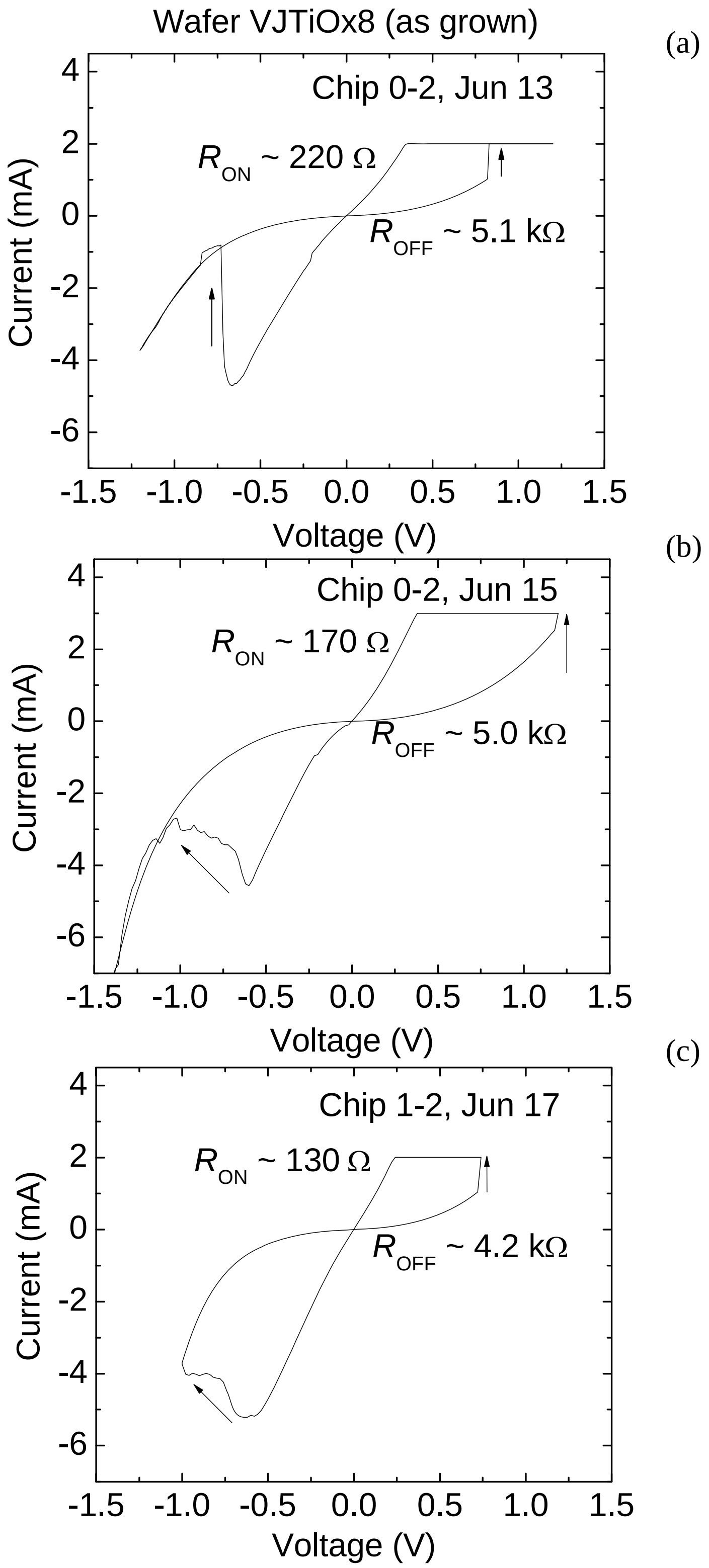

Fig. 6. DC *I-V* curves of three different devices from wafer VJTiOx8 before the RTA.

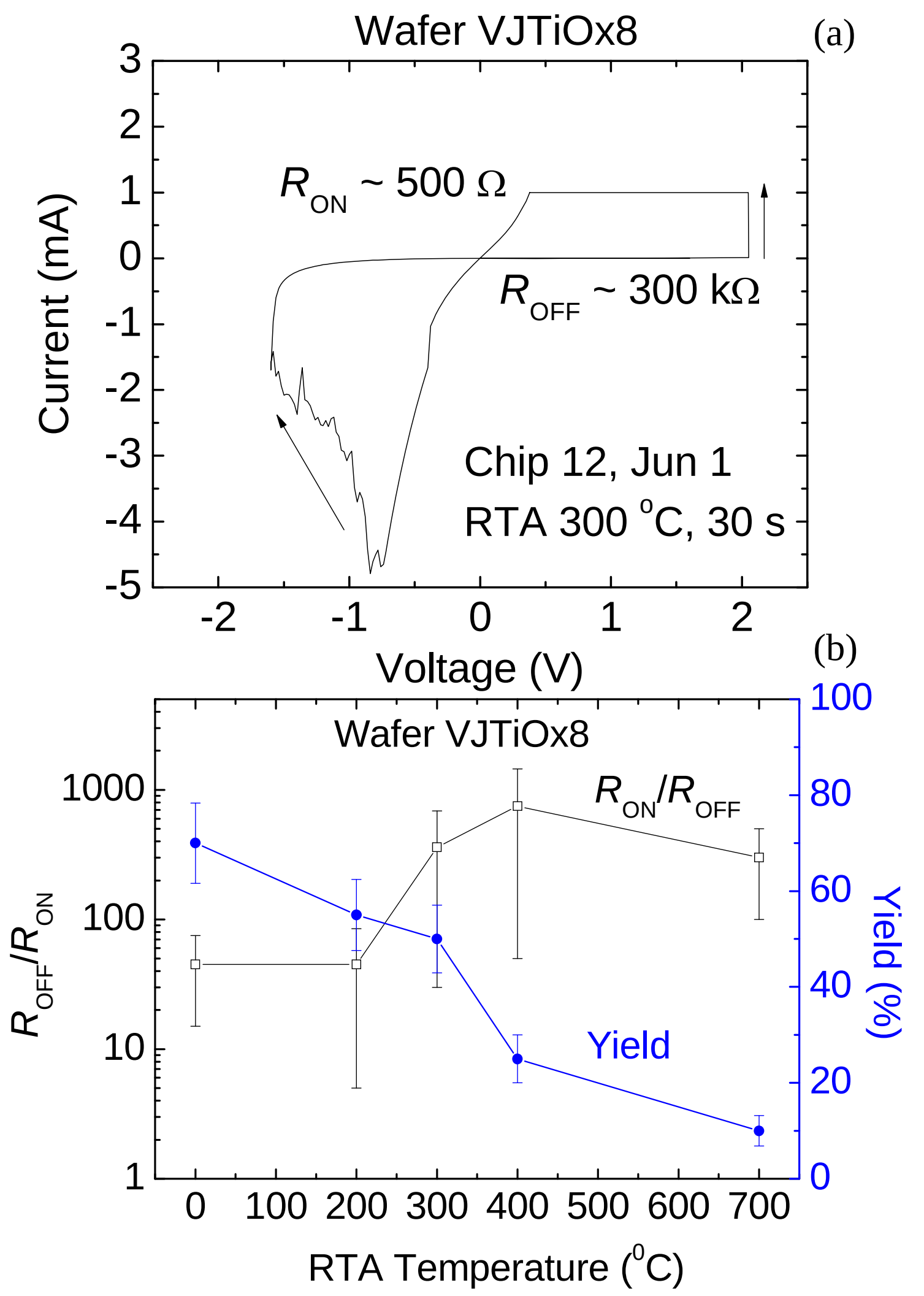

Fig. 7. (a) Typical dc *I-V* curve of a junction from wafer VJTiOx8 after the RTA, and (b) effect of temperature of a 30-second RTA on the OFF/ON resistance ratio and the yield of good devices from that wafer.

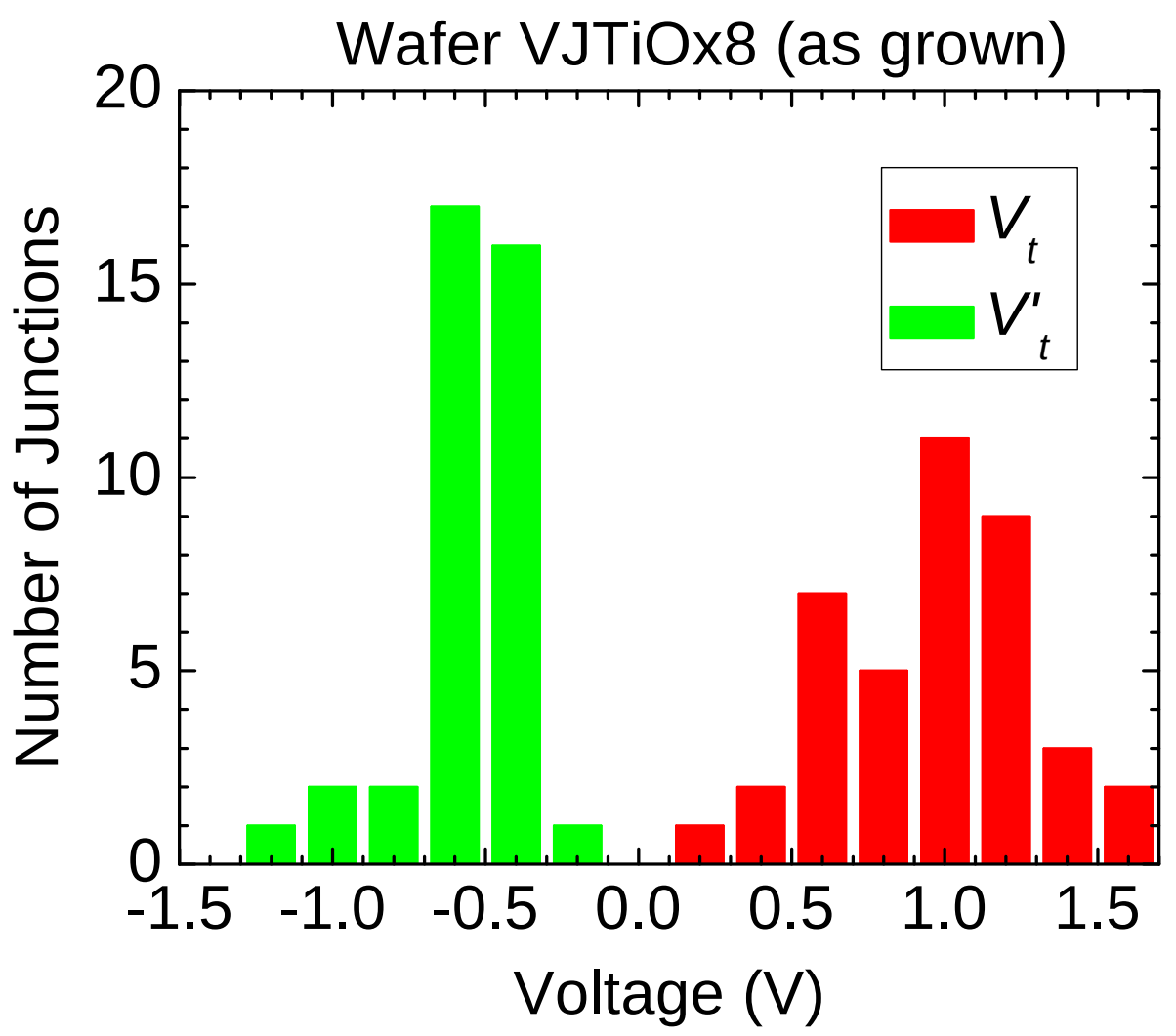

Fig. 8. Statistics of the switching thresholds $V_t$ and $V'_t$ for “as grown” devices (i.e. before the RTA) from wafer VJTiOx8.